\documentclass[preprint,showpacs,preprintnumbers,amsmath,amssymb]{revtex4}

\usepackage{graphicx}
\usepackage{dcolumn}
\usepackage{bm}

\begin{document}

\title{Test of Information Theory on the Boltzmann Equation}

\author{Kim Hyeon-Deuk \footnote{kim@yuragi.jinkan.kyoto-u.ac.jp}}

\affiliation{Graduate School of Human and Environmental Studies, Kyoto
University, Kyoto 606-8501, Japan}

\author{Hisao Hayakawa\footnote{hisao@yuragi.jinkan.kyoto-u.ac.jp}}

\affiliation{Department of Physics, Yoshida-South Campus, Kyoto University, Kyoto 606-8501, Japan}

\begin{abstract}
We examine information theory using the steady-state Boltzmann equation. 
In a nonequilibrium steady-state system under
steady heat conduction, the thermodynamic quantities from information
theory are calculated and compared with those
 from the steady-state Boltzmann equation. 
We have found that information theory is inconsistent with the steady-state Boltzmann equation. 
\end{abstract}



\maketitle

\newpage

\section{Introduction}\label{introduction}
The behaviors of gases in nonequilibrium states have received
considerable attention from the standpoint of understanding the
characteristics of nonequilibrium
phenomena. 
The Boltzmann equation is widely accepted as one of the most reliable
models for describing gases in nonequilibrium
phenomena, so that various attempts have been conducted on solving the Boltzmann equation.\cite{chapman,mac,resi,kogan,hand,kaper,cercignani1}  
Recently, we have derived the
explicit velocity distribution function of the steady-state Boltzmann
equation for hard-core molecules to second order in density and temperature gradients by the Chapman-Enskog method.\cite{kim} 

On the other hand, in the early 1960s, Zubarev\cite{zubarev,zubarev1} has developed nonequilibrium statistical mechanics and obtained the general form of a nonequilibrium velocity distribution
function with the aid of the maximum entropy principle. 
Thereafter the nonequilibrium velocity distribution
function to first order has been derived by expanding the Zubarev form
for the nonequilibrium velocity distribution function under some
constraints.\cite{katz} 

Jou and his coworkers have derived the nonequilibrium velocity distribution
function to second order by expanding the Zubarev form
for the nonequilibrium velocity distribution
function to second order under the some
constraints, which is called information
theory.\cite{jou,jou1,jou9,jou2,jou3,jou4}  
Information theory has attracted interest in the development of a general
framework for nonlinear nonequilibrium statistical mechanics. 
There is even a text book on information theory.\cite{jou} 
Jou \textit{et al.} have applied the velocity distribution function 
 from information theory to nonequilibrium dilute gases.\cite{jou,jou1,jou9,jou2,jou3,jou4,fort1,fort}  
There are also several applications of information theory to other microscopic theories, such as radiation\cite{jou5,jou6,jou7,jou8,fort3}, nonviscous
gases\cite{stella,stella1} and chemically reacting gases\cite{fort2}.  
Nettleton claimed that information theory
provides a statistical mechanical basis of irreversible processes 
and of extended thermodynamics which is consistent with the laws of
thermodynamics.\cite{net0} 
He has developed the maximum entropy formalism and applied it to a
dilute gas system.\cite{net,net1,net2,net3,eu}
However, in the actual applications, it is not easy to examine the validity of information 
theory.  
In order to demonstrate the invalidity of information theory, it is
necessary to find qualitative differences between information theory and 
the microscopic theories in the applications. 
In fact, though we have recently compared the effect of heat flux on the rate of
chemical reaction calculated from information theory with those which we have calculated from kinetic
theories, we have found no qualitative differences among them.\cite{kim1} 
We find no reports which conclude that information theory is not
an appropriate theory to describe nonequilibrium phenomena. 

However, as was mentioned in refs. 13 and 19, more examinations of
information theory should
be carried out from the microscopic
viewpoint to confirm whether there exists universality in nonlinear
nonequilibrium statistical mechanics. 
In the present paper, we check the validity of information
theory from a kinetic view point: we examine whether information
theory coincides with the steady-state Boltzmann equation, at least qualitatively, in a
nonequilibrium steady state. 

Suppose a dilute gas system subject to a temperature gradient along the
$x$-axis in a steady
state whose velocity distribution
function is expressed as $f=f(x,{\bf v})$. 
We introduce five conserved quantities and a heat flux playing important
roles in both information theory and the steady-state Boltzmann
equation. 
We define the density:  
\begin{equation}  
n(x)\equiv\int f d{\bf v}
,\label{bgk51}
\end{equation} 
and the temperature:  
\begin{equation}  
\frac{3n(x)\kappa T(x)}{2}\equiv
\int \frac{m{\bf v}^{2}}{2}f d{\bf v}
,\label{bgk52}
\end{equation}
with $m$ the mass of a molecule and $\kappa$ the Boltzmann constant. 
We assume no mean flow: 
\begin{equation} 
\int m {\bf v}f d{\bf v}={\bf 0}
,\label{bgk53}
\end{equation} 
where ${\bf 0}$ denotes the zero vector. 
Furthermore, we define the heat flux:  
\begin{equation}  
J_{x}\equiv\int \frac{m{\bf v}^{2}}{2}v_{x}f d{\bf v}. \label{bgk170} 
\end{equation}  
It should be emphasized that the heat flux $J_{x}$ calculated from
eq.(\ref{bgk170}) must be uniform in a steady state. 
Actually, in the case for the steady-state Boltzmann
equation, its solubility conditions lead to
the heat flux $J_{x}$ being constant to second order.\cite{kim} 

The organization of this paper is as follows. 
We will introduce information theory in $\S$ \ref{itsolve}. 
In $\S$ \ref{kt}, we will introduce the velocity distribution
functions of the steady-state Boltzmann equation for both hard-core and
Maxwell molecules to second
 order by the Chapman-Enskog method. 
In $\S$ \ref{application}, we will apply the velocity distribution
functions to a nonequilibrium steady-state system under steady heat conduction, and compare 
the results of thermodynamic quantities from information theory with those from the steady-state Boltzmann equation. 
Our discussion and conclusion are written in $\S$ \ref{discussion}. 

\section{Information Theory}\label{itsolve}
Let us introduce information theory proposed by Jou \textit{et
al.}\cite{jou,jou1,jou9,jou2,jou3,jou4}.  
The Zubarev form for the nonequilibrium velocity 
distribution function under a heat flux can be obtained by
maximizing the nonequilibrium entropy, defined as 
\begin{equation}  
S(x)\equiv-\kappa\int f\log f d{\bf v},\label{bgk130}
\end{equation}
under the constraints of the density (\ref{bgk51}), the temperature
(\ref{bgk52}), no mean flow (\ref{bgk53}) and the heat flux
(\ref{bgk170}) which is
now \textit{assumed} to be uniform as well as $n \kappa T$ by contrast
with the case for the steady-state Boltzmann
equation.\cite{kim}
Jou \textit{et al.} have finally obtained the nonequilibrium velocity 
distribution function to second order in the heat flux $J_{x}$ by
expanding the Zubarev's nonequilibrium velocity distribution
function to second order as
\begin{eqnarray}  
f=\frac{1}{Z}\exp\left(-\beta\frac{m{\bf v}^{2}}{2}\right)
\left[1-\frac{4J_{x}}{5n\kappa T}\left(\frac{m}{2\kappa T}\right)^{\frac{1}{2}}c_{x}\left(\frac{5}{2\beta\kappa T}-{\bf c}^{2}\right)
+\frac{4mJ_{x}^{2}}{25n^{2}\kappa^{3}T^{3}}c_{x}^{2}\left(\frac{5}{2\beta\kappa T}-{\bf c}^{2}\right)^{2}
\right],\nonumber \\
\label{bgk180} 
\end{eqnarray}
with the scaled velocity ${\bf c}\equiv (m/2\kappa T)^{1/2}{\bf v}$.     
Here $Z$ is given by 
\begin{equation}  
Z=\frac{1}{n}\left(\frac{2\pi}{\beta m}\right)^{\frac{3}{2}}\left(1+\frac{mJ_{x}^{2}}{5n^{2}\kappa^{3}T^{3}}\right),  
\label{bgk200} 
\end{equation}
in order to normalize $f$. 
The parameter $\beta$ is found to be  
\begin{equation}  
\beta=\frac{1}{\kappa T}\left(1+\frac{2mJ_{x}^{2}}{5n^{2}\kappa^{3}T^{3}}\right)\equiv\frac{1}{\kappa \theta},   
\label{bgk190} 
\end{equation}
and has been used by Jou \textit{et al.} to introduce $\theta$ as a
nonequilibrium temperature. 
From eq.(\ref{bgk190}) it is clear that the nonequilibrium temperature $\theta$ 
is not identical with the temperature $T$ defined 
in eq.(\ref{bgk52}), and $\theta$ is lower than $T$.\cite{jou,jou1,jou9,jou2,jou3,jou4,fort1,jou7,jou8}

By expanding the velocity 
distribution function (\ref{bgk180}) to second order in
$J_{x}$, we obtain the expression  for the \textit{modified} velocity 
distribution function: 
\begin{eqnarray}
f=
f^{(0)}\left\{1-\frac{3J_{x}}{2 n\kappa T}
\left(\frac{\pi m}{2\kappa T}\right)^{\frac{1}{2}}c_{x}
S_{\frac{3}{2}}^{1}({\bf c}^{2})
+\frac{2mJ_{x}^{2}}{5n^{2}\kappa^{3}T^{3}}
(1-{\bf c}^2)+
\frac{mJ_{x}^{2}}{5n^{2}\kappa^{3}T^{3}}c_{x}^{2}
[3\sqrt{\pi}S_{\frac{1}{2}}^{2}({\bf c}^{2})
+2]\right\},\nonumber \\ 
\label{bgk210} 
\end{eqnarray}  
with the local Maxwellian distribution function
$f^{(0)}=n(m/2\pi \kappa T)^{3/2}\exp(-{\bf c}^{2})$. 
Here $n$ and $T$ have been identified in eqs.(\ref{bgk51}) and
(\ref{bgk52}). 
This \textit{modified} velocity 
distribution function has been also obtained and used by Fort and Cukrowski\cite{fort1}. 
Note that $S_{k}^{p}(X)$ is the Sonine polynomial. (see, \textit{e.g.}
ref. 8) 
We have confirmed that the \textit{modified} velocity 
distribution function still satisfies constraints (\ref{bgk53}) and
(\ref{bgk170}), while the corrections appearing in eqs.(\ref{bgk200})
and (\ref{bgk190}) no longer appear in the density (\ref{bgk51}) and
the temperature (\ref{bgk52}) with 
the \textit{modified} velocity distribution function expressed in eq.(\ref{bgk210}). 
We adopt this \textit{modified} velocity distribution function instead
of the velocity distribution function shown in eq.(\ref{bgk180}) to
calculate macroscopic quantities in this paper. 
This adoption is based on the fact that the corrections in eqs.(\ref{bgk200})
and (\ref{bgk190}) are not significant, 
although Jou \textit{et al.}\cite{jou,jou1,jou9,jou2,jou3,jou4,fort1} believe
that the correction appearing in eq.(\ref{bgk190}) has important
physical meaning. (see also $\S$ \ref{discussion})  

\section{Kinetic Theory: the Steady-State Boltzmann Equation}\label{kt}
We introduce the velocity distribution function of the steady-state Boltzmann
equation for hard-core molecules which we
 have derived in ref. 8 valid to second
 order in density and the temperature gradient. 
In a nonequilibrium steady-state system under the
 temperature gradient along $x$-axis, it can be written as  
\begin{eqnarray}  
f=f^{(0)}\{1&-&\frac{4J_{x}}{5 b_{11} n\kappa T}(\frac{m}{2\kappa T})^{\frac{1}{2}}\sum_{r\ge 1}r! b_{1r}c_{x}\Gamma(r+\frac{5}{2}) S^{r}_{\frac{3}{2}}({\bf c}^{2})\nonumber\\
&+&\frac{4096mJ_{x}^{2}}{5625b_{11}^{2}n^{2}\kappa^{3}T^{3}}[\sum_{r\ge 2}r! b_{0r}\Gamma(r+\frac{3}{2})S^{r}_{\frac{1}{2}}({\bf c}^{2})\nonumber\\
&+&\sum_{r\ge 0}r! b_{2r}(2c_{x}^{2}-c_{y}^{2}-c_{z}^{2})\Gamma(r+\frac{7}{2})S^{r}_{\frac{5}{2}}({\bf c}^{2})]
\}, \label{be46}
\end{eqnarray}
where the specific values for $b_{1r}$, $b_{0r}$ and $b_{2r}$ are found in
Table \ref{bkr}. 
Note that we show only the values for $7$th Sonine approximation.\cite{kim} 

For our calculation of the
macroscopic quantities, we also adopt the precise velocity distribution function of the steady-state Boltzmann
equation for
Maxwell molecules to second order derived by Schamberg\cite{maxwell}. 
It becomes
\begin{eqnarray}  
f=f^{(0)}\{1&-&\frac{4J_{x}}{5 n\kappa T}(\frac{m}{2\kappa T})^{\frac{1}{2}}c_{x}\Gamma(\frac{7}{2}) S^{1}_{\frac{3}{2}}({\bf c}^{2})
+\frac{4096mJ_{x}^{2}}{5625 n^{2}\kappa^{3}T^{3}}[\sum_{r=2, 3}r! b_{0r}\Gamma(r+\frac{3}{2})S^{r}_{\frac{1}{2}}({\bf c}^{2})\nonumber\\
&+&\sum_{r=1, 2}r! b_{2r}(2c_{x}^{2}-c_{y}^{2}-c_{z}^{2})\Gamma(r+\frac{7}{2})S^{r}_{\frac{5}{2}}({\bf c}^{2})]
\},   
\label{be465}
\end{eqnarray}
where the precise values for $b_{0r}$ and $b_{2r}$ are written in
Table \ref{bkrmax}. 
It should be mentioned that the first-order velocity distribution
function in eq.(\ref{be465}) is identical with that for information theory shown in eq.(\ref{bgk210}), while the second-order velocity distribution functions are different from each
other. 

\section{Test of Information Theory}\label{application}
Now the velocity distribution functions to second order given in
eqs.(\ref{bgk210}), (\ref{be46}) and (\ref{be465}) shall be applied to the
nonequilibrium steady-state system. 
Note that all the definitions of physical quantities in this paper are the same as
those in ref. 8. 

To begin with, the pressure tensor in the nonequilibrium steady
state $P_{ij}$ becomes
\begin{eqnarray}  
P_{ij} &=& n\kappa T\left(\delta_{ij}+\lambda_{P}^{ij}\frac{mJ_{x}^{2}}{n^{2}\kappa^{3}T^{3}}\right),\label{bgk290}
\end{eqnarray}
with the unit tensor $\delta_{ij}$ and the numerical tensor components 
$\lambda_{P}^{ij}$ shown in Table \ref{macro1}. 
Note that the off-diagonal components of $\lambda_{P}^{ij}$
are zero and that $\lambda_{P}^{yy}=\lambda_{P}^{zz}$ is satisfied. 
We have found that $\lambda_{P}^{ij}$ for information theory is
qualitatively different from those for the steady-state Boltzmann
equation for both hard-core and Maxwell molecules: $P_{xx}$ becomes larger than $P_{yy}$ and $P_{zz}$ for information
theory\cite{jou9,jou2}, while $P_{xx}$ becomes smaller than $P_{yy}$ and $P_{zz}$ for the steady-state Boltzmann
equation for hard-core molecules, and no second-order corrections appear
in $P_{ij}$ for the steady-state Boltzmann equation for Maxwell molecules. 

Each component of the kinetic temperature in the nonequilibrium steady
state, i.e. $T_{\mathrm{i}}$ for $\mathrm{i}=x,y$ and $z$ is
also calculated as
\begin{eqnarray}  
\frac{n\kappa T_{\mathrm{i}}}{2} &=& \frac{n\kappa T}{2}\left(1+\lambda_{T_{\mathrm{i}}}\frac{mJ_{x}^{2}}{n^{2}\kappa^{3}T^{3}}\right),\label{bgk310}
\end{eqnarray} 
for $\mathrm{i}=x,y$ and $z$.  
Numerical values for the constants $\lambda_{T_{\mathrm{i}}}$ for $\mathrm{i}=x,y$ and $z$ are given in
Table \ref{macro1}. 
Note that $\lambda_{T_{y}}=\lambda_{T_{z}}$. 
We find that $\lambda_{T_{\mathrm{i}}}$ for information theory is qualitatively different from those for the steady-state Boltzmann
equation for both hard-core and Maxwell molecules: $T_{x}$ becomes larger than $T_{y}$ and $T_{z}$ for information
theory, while $T_{x}$ becomes smaller than $T_{y}$ and $T_{z}$ for the steady-state Boltzmann
equation for hard-core molecules, and no corrections appear in $T_{\mathrm{i}}$ for the steady-state Boltzmann equation for Maxwell
molecules. 

The Shannon entropy in the nonequilibrium steady
state becomes
\begin{eqnarray}  
S(x) &=& -n\kappa\log \left[n\left(\frac{m}{2\pi \kappa T}\right)^{\frac{3}{2}}\right]+\frac{3}{2}n\kappa+\lambda_{S}\frac{mJ_{x}^{2}}{n\kappa^{2}T^{3}},\label{bgk330}
\end{eqnarray}
to second order with the numerical constant $\lambda_{S}$ written in Table \ref{macro1}. 
It is found that $\lambda_{S}$ for information theory is identical to that obtained from the steady-state
Boltzmann equation for Maxwell molecules, while it is
slightly different from $\lambda_{S}$ calculated from the steady-state
Boltzmann equation for hard-core molecules. 
This is because the correction term for the Shannon entropy is determined
only by the first-order
velocity distribution function, as was indicated in ref. 12. 

\section{Discussion and Conclusion}\label{discussion}
It is seen that the first-order velocity distribution
 functions for the steady-state
Boltzmann equation for both hard-core and
Maxwell molecules, i.e. the first-order terms in eqs.(\ref{be46}) and (\ref{be465}), are
consistent with that derived by expanding Zubarev's velocity distribution
function\cite{zubarev,zubarev1,katz}. 
This consistency is attributed to the fact that a nonequilibrium correction in the nonequilibrium entropy should 
 appear to even order of a nonequilibrium flux, \textit{e.g.}
 $\delta S\propto -J_{x}^{2}$, in order that the nonequilibrium entropy has a
 maximum at $J_{x}=0$, and that a thermodynamic force $F=\partial \delta S/\partial J_{x}$  which drives a nonequilibrium system towards the
 state of equilibrium is proportional to the nonequilibrium
 flux.\cite{onsager}
This fact leads to a conclusion that the nonequilibrium entropy is not modified from the local
equilibrium entropy to first order, and that the Shannon-type entropy is appropriate as
 the nonequilibrium entropy to first order. 

On the other hand, we have confirmed that both forms (\ref{be46}) and
(\ref{be465}) of the second-order 
velocity distribution functions differ from that suggested by
information theory\cite{jou,jou1,jou9,jou2,jou3,jou4}. 
Although Jou \textit{et al.} have applied 
information theory to nonequilibrium dilute
gases\cite{jou,jou1,jou9,jou2,jou3,jou4,fort1,fort}, we have found that information theory contradicts the steady-state
Boltzmann equation: all the macroscopic quantities for information
theory except for the
Shannon entropy $S$ in eq.(\ref{bgk330}) are qualitatively different from those for the steady-state
Boltzmann equation for both hard-core and Maxwell molecules. 
These qualitative differences between information theory
 and the steady-state
Boltzmann equation still appear no matter which boundary
 condition is adopted. 
It is conjectured that the entropy defined in eq.(\ref{bgk130})
is not appropriate as the nonequilibrium entropy to second order, though the
Shannon-type entropy has 
been widely used as the nonequilibrium entropy to any order.\cite{zubarev,zubarev1,katz,jou,jou1,net0,net,net1,net2,net3,eu,onsager} 
We emphasize that it is probably the first time to find qualitative
differences between information theory and nonequilibrium microscopic theories and demonstrate that information theory is
inconsistent with the nonequilibrium microscopic theories. 
We can conclude that, though quite a few statistical physicists have believed the existence of a universal velocity distribution
function in the nonequilibrium steady state by maximizing the
Shannon-type
entropy\cite{zubarev,zubarev1,katz,jou,jou1,net0,net,net1,net2,net3,eu},
the universal velocity distribution
function does not exist in the nonequilibrium steady state. 
It is also worth mentioning that, although information theory based
on the Tsallis entropy has been also developed\cite{tsallis,tsallis1}, 
the general form of the velocity distribution function for information theory based
on the Tsallis entropy\cite{tsallis,tsallis1} cannot be expanded even to first order because the expanded velocity
distribution function diverges. 

We have also confirmed that, in all the macroscopic quantities calculated
in the present paper, there are no differences between the results from the \textit{modified} velocity 
distribution function given in eq.(\ref{bgk210}) and those from Jou's velocity 
distribution function shown in eq.(\ref{bgk180}) so long as the same boundary
condition is adopted. 
This suggests that the nonequilibrium temperature $\theta$ has no physical significance. 
We emphasize that the identifications of the density, the
temperature and the mean flow ( see eqs.(11), (12) and (13) in ref. 8) do not affect the physical properties of the velocity
distribution function for the steady-state Boltzmann equation\cite{kim},  
and that those identifications must be satisfied for the conservation laws 
in the case for the
steady-state Bhatnagar-Gross-Krook(BGK) equation.\cite{santos,santos1}

\begin{acknowledgments}
We would like to express our sincere thanks to H. Tasaki who made us aware of the significance
of understanding nonequilibrium steady-state phenomena. 
This research was essentially inspired by him. 
We are grateful to S. Sasa who has always
had crucial, interesting and cheerful discussions with us and has encouraged us to carry out these calculations. 
The authors also appreciate Ooshida T., A. Yoshimori, 
M. Sano, J. Wakou, K. Sato, H. Kuninaka, T. Mizuguti, T. Chawanya, S. Takesue 
and H. Tomita for fruitful discussions and useful comments. 
This study has been supported partially by the Hosokawa Powder Technology
Foundation, the Inamori Foundation and Grant-in-Aid for Scientific Research (No. 13308021).  
\end{acknowledgments}

\newpage

\begin{center}
\begin{table}[h]
\begin{ruledtabular}
\caption{\label{bkr}Numerical constants $b_{1r}$, $b_{0r}$ and $b_{2r}$ in
 eq.(\ref{be46}). 
All the values are ones for $7$th Sonine approximation.
 }
\begin{tabular}{@{\hspace{\tabcolsep}\extracolsep{\fill}}cccc} 
{}&{$b_{1r}$}&{$b_{0r}$}&{$b_{2r}$}\\ \hline
{$0$}&{$0$}&{$1$}&{$-3.320\times 10^{-2}$} \\ \hline
{$1$}&{$1.025$}&{$0$}&{$-1.276\times 10^{-1}$} \\ \hline
{$2$}&{$4.892\times 10^{-2}$}&{$4.380\times 10^{-1}$}&{$6.414\times 10^{-2}$} \\ \hline
{$3$}&{$3.715\times 10^{-3}$}&{$-5.429\times 10^{-2}$}&{$5.521\times 10^{-3}$} \\ \hline
{$4$}&{$2.922\times 10^{-4}$}&{$-4.098\times 10^{-3}$}&{$4.214\times 10^{-4}$} \\ \hline
{$5$}&{$2.187\times 10^{-5}$}&{$-3.184\times 10^{-4}$}&{$3.106\times 10^{-5}$} \\ \hline
{$6$}&{$1.492\times 10^{-6}$}&{$-2.087\times 10^{-5}$}&{$1.861\times 10^{-6}$} \\ \hline
{$7$}&{$8.322\times 10^{-8}$}&{$-$}&{$-$} \\ 
\end{tabular}
\end{ruledtabular}
\end{table}
\end{center}

\begin{center}
\begin{table}[h]
\begin{ruledtabular}
\caption{\label{bkrmax}Numerical constants $b_{0r}$ and $b_{2r}$ in
 eq.(\ref{be465}). }
\begin{tabular}{@{\hspace{\tabcolsep}\extracolsep{\fill}}ccc} 
{}&{$b_{0r}$}&{$b_{2r}$}\\ \hline
{$1$}&{$-$}&{$\frac{75}{896}$} \\ \hline
{$2$}&{$\frac{825}{1024}$}&{$\frac{125}{1536}$} \\ \hline
{$3$}&{$-\frac{25}{256}$}&{$-$} \\ 
\end{tabular}
\end{ruledtabular}
\end{table}
\end{center}

\begin{center}
\begin{table}[h]
\begin{ruledtabular}
\caption{\label{macro1}The numerical constants for the macroscopic
 quantities: the precise values for information theory, 
the $7$th Sonine approximation values for hard-core molecules and the exact values for Maxwell molecules. 
 }
\begin{tabular}{@{\hspace{\tabcolsep}\extracolsep{\fill}}cccccc} 
{}&{$\lambda_{P}^{xx}$}&{$\lambda_{P}^{yy}$}&{$\lambda_{T_{x}}$}&{$\lambda_{T_{y}}$}&{$\lambda_{S}$} \\ \hline
{information theory}&{$\frac{12}{25}$}&{$-\frac{6}{25}$}&{$\frac{6}{25}$}&{$-\frac{3}{25}$}&{$-\frac{1}{5}$} \\ \hline 
{hard-core molecules}&{$-4.600\times 10^{-2}$}&{$2.300\times 10^{-2}$}&{$-2.300\times 10^{-2}$}&{$1.150\times 10^{-2}$}&{$-2.035\times 10^{-1}$} \\ \hline
{Maxwell molecules}&{$0$}&{$0$}&{$0$}&{$0$}&{$-\frac{1}{5}$}\\ 
\end{tabular}
\end{ruledtabular}
\end{table}
\end{center}

\end{document}